\def\Version{ 2.00
    }

  %%%%%%%%%%%%%%%%%%%%%%%%%%%%%%%%%%%%%%%%%%%%%%%%%%%
  %%%  PROCESS THIS FILE WITH `tex' and `dvips'  %%%%
  %%%%%%%%%%%%%%%%%%%%%%%%%%%%%%%%%%%%%%%%%%%%%%%%%%%

%% \headline{\hfil {{\bf DRAFT} {\it Not for distribution} \  \Version \ {\bf DRAFT}} \hfil}

% Outlining is at end!

%: Page Setup for 8.5 x 11 inch paper        

\message{ Assuming 8.5" x 11" paper }    

\magnification=\magstep1	          % \magstep1=1200

\raggedbottom

\overfullrule=0pt % to delete the black boxes marking overfull hbox

\parskip=9pt

% \hsize=6.4 true in
% \vsize=8.7 true in
%
% \hoffset=0.27 true in
% \voffset=0.28 true in

\def\singlespace{\baselineskip=12pt}      % spacing for stuff like abstract
\def\sesquispace{\baselineskip=16pt}      % spacing for main text
% \def\sesquispace{\baselineskip=20pt}    %%% draft spacing %%%

%: Load or define some TeX macros            
%:  general purpose macro files              

%% \input eplain   % [not installed on mars apparently]

%% Incorporate these files here, when paper finished
% \input mathmacros
\font\openface=msbm10 at10pt
\def\Integers      {{\hbox{\openface Z}}}
\def\ideq{\equiv}		% triple equal sign
\def\braces#1{ \{ #1 \} }

\def\PrintVersionNumber{
 \vskip -1 true in \medskip 
 \rightline{version \Version} 
 \vskip 0.3 true in \bigskip \bigskip}

\def\linebreak{\hfil\break}
\def\lbr{\linebreak}

\font\titlefont=cmb10 scaled\magstep2 
\def\author#1 {\medskip\centerline{\it #1}\bigskip}
\def\address#1{\centerline{\it #1}\smallskip}

\def\furtheraddress#1{\centerline{\it and}\smallskip\centerline{\it #1}\smallskip}

\def\email#1{\smallskip\centerline{\it address for email: #1}} 

\def\AbstractBegins
{
 \singlespace                                        % spacing for abstract
 \bigskip\leftskip=1.5truecm\rightskip=1.5truecm     % begin indentation
 \centerline{\bf Abstract}
 \smallskip
 \noindent	% this doesn't seem to take effect over a blank line
 } 
\def\AbstractEnds
{
 \bigskip\leftskip=0truecm\rightskip=0truecm       %  end indentation
 }

\font\headingfont=cmb10 at 12pt
\font\subheadfont=cmssi10 scaled\magstep1 % sans serif italic
\def\section #1 {\bigskip\noindent{\headingfont #1 }\par\nobreak\smallskip\noindent}
\def\subsection #1 {\medskip\noindent{\subheadfont #1 }\par\nobreak\smallskip\noindent}

\def\ReferencesBegin
{
 \singlespace					   % single spacing
 \vskip 0.5truein
 \centerline           {\bf References}
 \par\nobreak
 \medskip
 \noindent
 \parindent=2pt
 \parskip=6pt			% earlier was 10 pt and then 4pt
 }

\def\reference{\hangindent=1pc\hangafter=1} 
\def\ref{\reference}

\def\journaldata#1#2#3#4{{\it #1\/}\phantom{--}{\bf #2$\,$:} $\!$#3 (#4)}
\def\eprint#1{{\tt #1}}
\def\arxiv#1{\hbox{\tt http://arXiv.org/abs/#1}} 

%:  possibly load { miniltx } and/or { graphicx } here 

\input graphicx                 % this inputs miniltx

% \input miniltx

%:  macro for figure caption (EXPERIMENTAL)  

\def\Caption#1{\vbox{

 \leftskip=1.5truecm\rightskip=1.5truecm     % begin indentation
 \singlespace                                % begin single spacing
 \noindent #1
 \vskip .25in\leftskip=0truecm\rightskip=0truecm}
 \sesquispace}
 %
 % It might be significant that the second line is blank!

%:  ad hoc macros for this paper             

\def\omegabar{\overline\omega}

%: phantom input	   	       	     

% Not clear why, but following somehow helps with layout of first page:

\phantom{}

%: Possibly print a version number           

\PrintVersionNumber   % [[ comment out? ]]
% \PrintTimestamp       % comment out? 

%: Title                                     

\sesquispace
\centerline{{\titlefont Does a Quantum Particle Know its Own Energy?}\footnote{$^{^{\displaystyle\star}}$}%
{To appear in {\it Journal of Physics: Conf. Ser.}
 %% \journaldata{Journal of Physics: Conf. Ser.}{}{}{}, [[insert when known]]
 \lbr 
%% \eprint{http://arxiv.org/abs/@@@@}  [[insert when known]]
}}

\bigskip

%: Authors                                   

\singlespace			        % (spacing for addresses etc.)

\author{Rafael D. Sorkin}
\address
 {Perimeter Institute, 31 Caroline Street North, Waterloo ON, N2L 2Y5 Canada}
\furtheraddress
 {Department of Physics, Syracuse University, Syracuse, NY 13244-1130, U.S.A.}
\email{rsorkin@perimeterinstitute.ca}

\AbstractBegins                              
If a wave function does not describe microscopic reality then
what does?  Reformulating quantum mechanics in path-integral
terms leads to a notion of ``precluded event" and thence to the
proposal that quantal reality differs from classical reality in
the same way as a set of worldlines differs from a single
worldline.  One can then ask, for example, which sets of
electron trajectories correspond to a Hydrogen atom in its
ground state and how they differ from those of an excited state.
We address the analogous questions for simple model that
replaces the electron by a particle hopping (in discrete time)
on a circular lattice.

\AbstractEnds                                

\bigskip

%: Body of paper                             

%% Turn on spacing for body of paper

\sesquispace
\vskip -10pt

\section{} % Untitled Introductory section 
Should we try to form for ourselves an image of the quantum world?  Or
must our theories find their meaning solely in assertions about
laboratory instruments and their readings? 
In other words, is it permissible to ask the question, 
{\it To what `reality' does the quantum formalism refer?}
I believe that this question is not only a legitimate one, but it is one
that we must ask.  If we try to avoid it, we fall into a vicious circle
because instruments are made of atoms and not vice versa.  More
importantly, quantum gravity and, especially, cosmology need to deal
with parts of nature where one finds neither observers nor instruments,
and to which an ``operational point of view'' therefore seems unsuited.

Beyond the more or less familiar reasons just adduced to support the
claim that the problems of quantum gravity and ``quantum foundations''
are intertwined, there's another connection that could be important
via the concept of 
relativistic causality.  
If the condition
that ``physical influences propagate causally'' could be given an
intrinsic formulation, free of references to external observers, and if
the resulting criterion were formulated in terms of histories (as Bell's
``local causality'' is, for example), then one should be able to decide
whether or not quantum mechanics and quantum field theory 
satisfied 
this
condition.  If they did, then it would make sense to require it also of
quantum gravity, and this in turn could be the key to constructing a
viable quantum dynamics for causal sets.

Perhaps the challenge of ``quantum foundations" is not so urgent for
quantum computing, which is concerned more with the manipulation of
information than with microscopic reality as such.  But even
there, it seems possible that a more definite picture of the micro-world
could someday lead 
us to widen our
conceptions of measurement and of
% what constitutes a 
computation.

What I hope to illustrate 
in this paper
is a possible answer to the 
italicized
question
posed above, an answer that arose from thinking of quantum mechanics in
the language of histories, i.e. the language of the path integral.
According to this answer, the micro-world is described by something
called a {\it coevent\/}, but instead of attempting to define that
concept in the general case, I will consider a very simple model of a
particle hopping on a lattice, and in that setting, will present a
short calculation that I hope will indicate more concretely how the
proposal is meant to go.
In the particular scheme I will present, reality will be something like
a trajectory or worldline of the hopping particle, but instead of being
a single trajectory, as in classical physics, it will be a set of
trajectories.  This particular choice is not necessarily the best one,
but it is the simplest and therefore appropriate to illustrate the main
idea.

Notice that in a coevent formulation, reality is {\it not} represented
by a wave function $\psi$.  Therefore, although a concept like position
will have a straightforward meaning, a concept like momentum or energy
will not.  But then we have to ask how an electron in a hydrogen atom
``knows'' for example, whether the atom is in its ground state or in an
excited state.  Is there something about the trajectories that carries
this information?  Hence the title of this paper.

\section{I.~ A simple unitary model --- the n-site hopper} 
A simple model allowing one to refer to trajectories and ground/excited
states is the ``$n$-site hopper'', by which I mean a particle residing
on an $n$-site periodic lattice, and at each of a discrete succession of
moments either staying where it is or jumping to some other site, the
respective amplitudes being those given by 
the simple ``transfer matrix'' reproduced below.~[1][2]\ 
In this ``toy world'', nothing exists beyond
the hopper itself, and since both space and time are discrete, the
possible realities or {\it coevents} can be computed with minimal
difficulty.  

In order to describe the hopping amplitudes, let us identify the nodes
of the lattice with 
the elements of $\Integers_n$,
the integers modulo $n$.  
Further
let $x\in\Integers_n$ be the location of the particle at some moment
and let $x'$ be its location at the next moment, 
and write for brevity $\exp(2\pi i z) \ideq {\bf 1}^{z}$.
The amplitude to go from $x$ to $x'$ in a single step is then
$$
     {1 \over \sqrt{n}} \ {\bf 1}^{ (x-x')^2 / n}         \eqno(1a)
$$
for $n$ odd, 
and 
$$
    {1 \over \sqrt{n}} \ {\bf 1}^{(x-x')^2 / 2n}          \eqno(1b)
$$
for $n$ even.  For example, for $n=6$ and with $q={\bf 1}^{1/12}$, the 
(un-normalized) amplitudes to hop by 0, 1, 2 or 3 sites respectively are
$q^0=1$, $q^1=q$, $q^4$, and $q^9=-i$. 

It is not difficult to verify that these amplitudes (more precisely the
matrix they comprise) are unitary.  Interestingly, they take precisely
the form of the propagator of a non-relativistic free particle in
one-dimension, suggesting that in a suitable $n\to\infty$ limit, this
hopper model could provide a fully self-consistent regularization of the
path-integral for such a particle.

For the 2- and 3-site hoppers, the amplitudes are particularly simple,
yielding for $n=3$ the matrix
$$
    {1 \over \sqrt{3}} \pmatrix{1 & \omega & \omega \cr 
                               \omega & 1 & \omega \cr 
                               \omega & \omega & 1 \cr } 
    \qquad\qquad (\omega={\bf 1}^{1/3})
$$
and for $n=2$ the  matrix
$$
             {1 \over \sqrt{2}} \pmatrix{1 & i \cr i & 1} \ .   
$$

\section{II.~ {Review of anhomomorphic coevents and the ``Multiplicative Scheme''}}  
In the formulation I am advocating,
``nature''
is represented 
in terms of {\it histories}, which for present purposes means
trajectories of the hopping 
particle.~[3][4][5]\ 
Physical reality --- a ``possible world'' --- is then described by a
{\it coevent\/} which specifies 
which events happen and which don't, 
where an {\it  event\/} is by definition a set of histories.\footnote{$^\star$}
{not to be confused with the word ``event'' used to denote a point of
 spacetime.  The definition here follows the usage in probability
 theory, which in turn is more in tune with the everyday meaning of the
 word.}
More formally, a coevent will be a function $\phi$ that assigns either
$0$ or $1$ to each event, according as the event doesn't or does happen
in the world described by $\phi$.

From this perspective, our description of the physical world would be
complete if we were able to specify fully the actual coevent $\phi$.
The role of ``dynamics'', then, is to help us toward a fuller such
specification by placing conditions on $\phi$ that narrow down the range
of possibilities that need to be considered.  I will assume that the
input to this dynamics takes the form of a path integral (or in the case
of the hopper, a path sum).  Wave functions $\psi$, insofar as they play
a role at all, will provide provisional initial amplitudes that go into
the computation of the path-sum (as an approximate summary of past).

But is it possible to base a dynamics on a path-sum alone?  If we wish
to do so then, plainly, we must construe the latter as something more
than a technical device to compute transition amplitudes between some
initial wave-function and some final one.  Instead we will interpret it
as providing for any event $A$, the {\it quantal measure\/} {$\mu(A)$} of
that event.\footnote{$^\dagger$}
{$\mu(A)$ is also known as the diagonal element of the decoherence functional $D(A,A)$.}
Once again, I will omit the formal 
definition [6][7][8],
since we will see very
soon how concretely to compute {$\mu(A)$} in the case at hand.

\subsection {The preclusion principle}  
It is not hard to convince oneself that in one special case $\mu(A)$ has
the meaning of an ordinary probability, namely when the event $A$ can be
described as a possible outcome (``instrument reading'') of a given
laboratory experiment.  In such a case, the analyses of numerous
gedankenexperiments over the years have made it plausible (albeit people
don't always express it this way) that $\mu(A)$ coincides with the
Born-rule probability of the 
outcome $A$.  
%% outcome $A$.~[R::born-rule]\  [[reinstate if find ref]]
But if this is accepted, then
it follows immediately that a zero value of $\mu(A)$ implies that the
corresponding instrument-event almost surely does not occur --- it is
{\it precluded\/}.\footnote{$^\flat$}
{To this extent, $\mu(A)$ is a sort of ``propensity'' for the event
 $A$ to happen.  One cannot go further and interpret it as a genuine
 probability, because, thanks to quantal interference, it is not
 additive on disjoint events.} 
Extending this conclusion to the
case of an arbitrary --- macroscopic or microscopic --- event, we
arrive at a dynamical principle of general applicability:
  {\it If $\mu(A)=0$ then the event $A$ cannot happen.}

% if the path-integral defining $\mu$ is to tell us anything at all about
% the ``propensity'' for events more general than ``pointer readings'',
% then it's hard to see how one could give up the conclusion that
% $\mu(A)=0$ entails preclusion.  
 
\subsection {Preclusion and primitivity}  
The preclusion principle requires of any dynamically viable coevent
$\phi$ that it deny every event whose quantal measure vanishes:
$\phi(A)=0$ if $\mu(A)=0$.
We have seen that this principle flows naturally from the path-integral,
but by itself it is still rather weak, in the sense that a vast number
of coevents can satisfy it even when preclusions abound.  In our hopper
example, for instance, we will have 27 histories, and consequently 
$2^{2^{27}}=2^{134217728}$ coevents in toto.  The number of precluded
events is also large (2017807), but even so, there remain
$2^{132199921}$ coevents which are {\it preclusive} in the sense that
they satisfy our condition.  On the other hand, $2^{132199921}$ is only
a tiny fraction of $2^{134217728}$, so one might feel on the contrary
that the preclusion principle is rather strong.

Be that as it may, I think the weightiest reason why the preclusion
principle cannot stand alone is that it seems incapable of yielding the
classical conception of reality-as-a-single-history when the measure
$\mu$ is classical 
(and possibly  ``deterministic'').  
Thus, preclusion alone would
not exclude coevents for which an experiment had more than one
macroscopic outcome,
or for which no definite outcome at all happened.  In other words it
would not resolve the ``measurement problem''.

To complete the dynamical story, then, we will supplement preclusion
with a further principle of ``minimality'' or {\it primitivity\/} designed
to remedy the deficiencies just cited.  
How properly to frame such a supplementary condition is a question
not yet settled, but the simplest proposal is that of the so-called 
{\it multiplicative scheme\/}, and this is the one I will adopt
for the present analysis.

%:  figure 1

\vbox{
   \bigskip

  \includegraphics[scale=0.5]{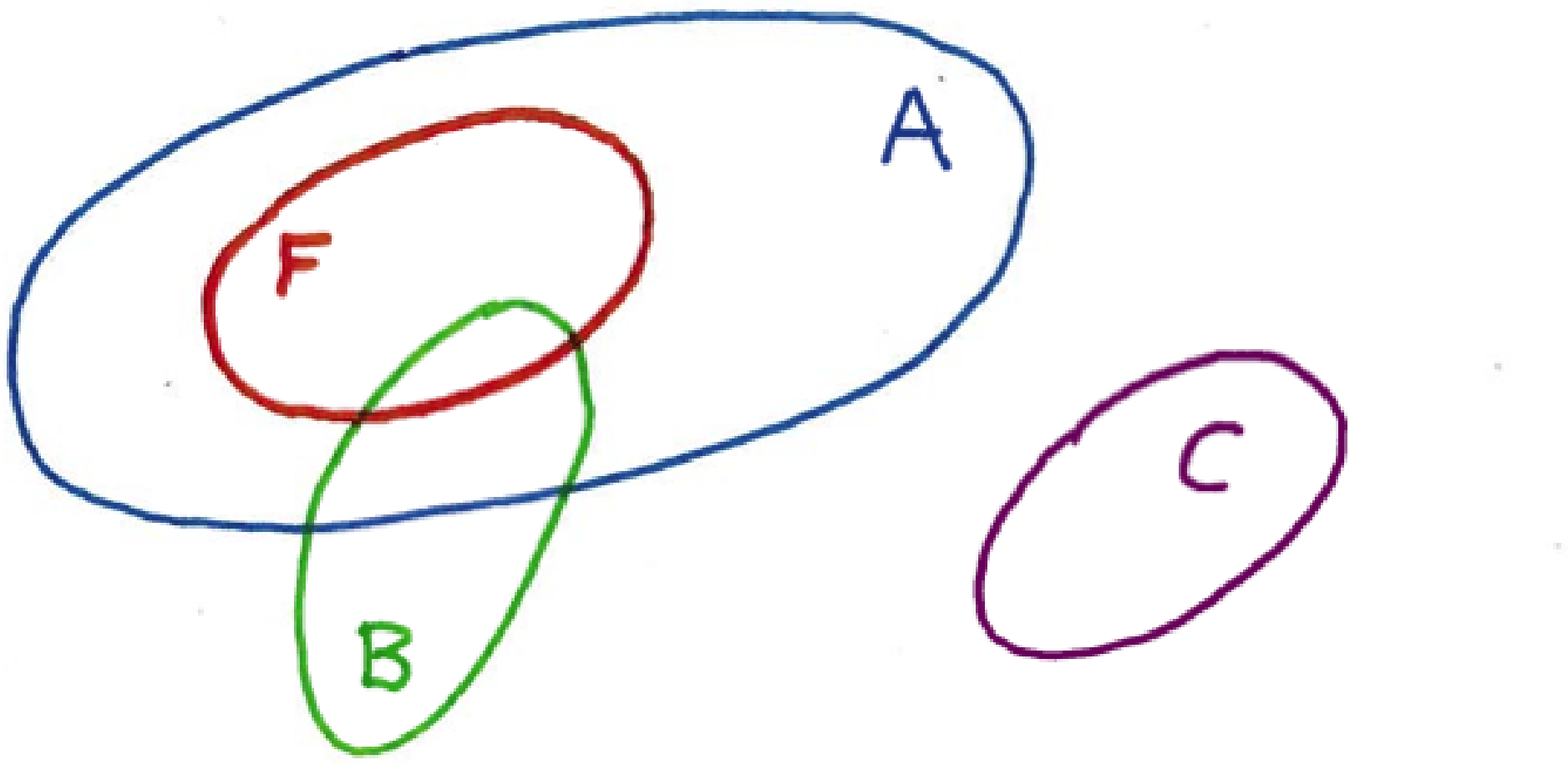} %%%  

  \Caption{{\it Figure 1.} Three events and a multiplicative coevent.
   The three events are 
   %% represented by 
   three sets of histories, $A$, $B$, $C$, while
   the coevent $\phi$ 
   corresponds to
   %% is represented by 
   a further set of histories $F$
   called its {\it support\/}.
   In the ``reality'' described by $\phi$, $A$
   happens, while $B$ and $C$ do not happen.  
   In formulas, $\phi=F^*$, $\phi(A)=1$, and $\phi(B)=\phi(C)=0\,$.
   }}

\subsection {the multiplicative scheme}  
Classical physics identified reality with a single history, but that no
longer seems possible quantum mechanically because the characteristic
phenomenon of {\it interference} produces non-classical patterns of
preclusion which seem to demand a modified conception of reality.

In place of a single history, the multiplicative scheme describes the
physical world by a coevent of the form $\phi=F^*$, where $F$ is now a
{\it set\/} of histories that reduces to a singleton set only in very
special (effectively classical) circumstances.  
As illustrated in figure 1, 
the coevent $F^*$ assigns 1 (`true') to an event $A$ if and
only if $A$ is a superset of $F$.  
Thus in the diagram, if $\phi=F^*$ then $\phi(A)=1$ while
$\phi(B)=\phi(C)=0\,$.  When $\phi=F^*$ I will refer to $F$ as the
{\it{}support} of $\phi\,$.  Within the multiplicative scheme, a coevent
is thus fully determined by its support.\footnote{$^\star$}
{A little thought should convince you that 
% the same rule 
 the rule that $\phi(A)=1$ iff $F\subseteq A$ also
 holds
 classically, with $F$ being the set that contains the ``actual
 history'' as its sole element. }

Now when is a coevent $\phi=F^*$ ``dynamically viable'' within the
multiplicative scheme?  By assumption it must be preclusive, and this
means precisely that its support $F$ must not fall wholly within any
precluded event.  Beyond this, we require further that $F$ be as
small as possible consistent with the preclusivity condition just
stated.  A coevent (or its support) that fulfills all these conditions I
will call {\it primitive preclusive}, or for short just {\it primitive}.
A primitive coevent thus describes a ``possible world'', where
``possible'' is to be understood relative to the given set of preclusions.

One sees immediately that in the absence of any preclusions (other than
the empty event itself) a primitive preclusive support will consist
solely of a single history, and the same holds whenever the pattern of
preclusions is of the type that occurs in either classical deterministic,
or classically stochastic theories. 
Much more than this could be said about the multiplicative scheme and
its consequences,~[9][10][11]\ 
but now I want to focus on a very concrete example and
on our specific question: {Does the particle know its own energy?}

\section{III.~ Histories and amplitudes for the 3-site hopper}           
In simple cases, preclusion is decided by whether the sum of the
amplitudes vanishes.  This will be true for our example, and it will
make it easy to find the primitive supports.

To simplify as much as possible, let the hopper take three steps and
then stop.  And let there be nothing else in the world beside this hopper.
Our space of histories then comprises exactly 81 trajectories, depending
on where the hopper starts from and 
where it lands
% how it moves 
at each of the three
subsequent moments.  
In fact, however, we only need to consider the 27 histories shown in figure 2, 
because one can prove, 
as a general feature of the multiplicative scheme, that 
any primitive support must correspond to 
a sharp final position.  
%% a sharp final position.~[R::precl-sep]\ [[reinstate when found or published]]
That is, one of the three events,
``the hopper terminates at 0'',
``the hopper terminates at 1'', or
``the hopper terminates at 2'',
must happen.
Without loss of generality we can suppose it is the first of these
events, and this is what the figure illustrates.

With the final position fixed at $x=0$, it is also very easy to decide
whether a given set of histories is precluded: this occurs iff the
amplitudes of the constituent histories sum to zero.  To work out the
pattern of preclusions, it thus suffices to know all the amplitudes.

Now given a history $\gamma$, its net amplitude is the amplitude it
inherits from its starting location, multiplied by the amplitudes of the
individual hops, the former being given by what one might call the
``initial wave-function'' $\psi_{initial}$.
For $\psi_{initial}$ let us consider two possible choices, $\psi_{0}$
and $\psi_{+}$, which we may
call by way of analogy ``ground state'' and ``traveling wave''.  
The amplitudes for these are respectively
$(\psi(0),\psi(1),\psi(2))=(1,1,1)$
and
$(\psi(0),\psi(1),\psi(2))=(1,\omega,\omega^2)$,
where
$\omega={\bf 1}^{1/3}$ as before.
Clearly both 
$\psi_{0}$ and $\psi_{+}$
% of these $\psi$'s 
are eigenvectors of the ``transfer
matrix'' defined earlier.
(The overall normalization of $\psi_{initial}$ is immaterial since it
has no effect on preclusion.) 

It is now straightforward to compute the amplitudes of our 27 histories
in each case.  The figure exhibits them for the traveling wave, and
those for the ground-state are similar (but even simpler to compute
since the dependence on the starting position is absent.)  Curiously,
the amplitudes per se (though not of course the way they are distributed
among the histories) turn out to be exactly the same for both cases.
Counting them up, one obtains the following multiset:
$$
   { \omegabar^{\braces{12}} \qquad 1^{\braces{9}} \qquad  \omega^{\braces{6}} }\ ,
$$
meaning 12 histories with amplitude $\omega^2=\omegabar$, 9 with amplitude 1, and
6 with amplitude $\omega$.

%:  figure 2
\vbox{
   \bigskip

  \includegraphics[scale=0.5]{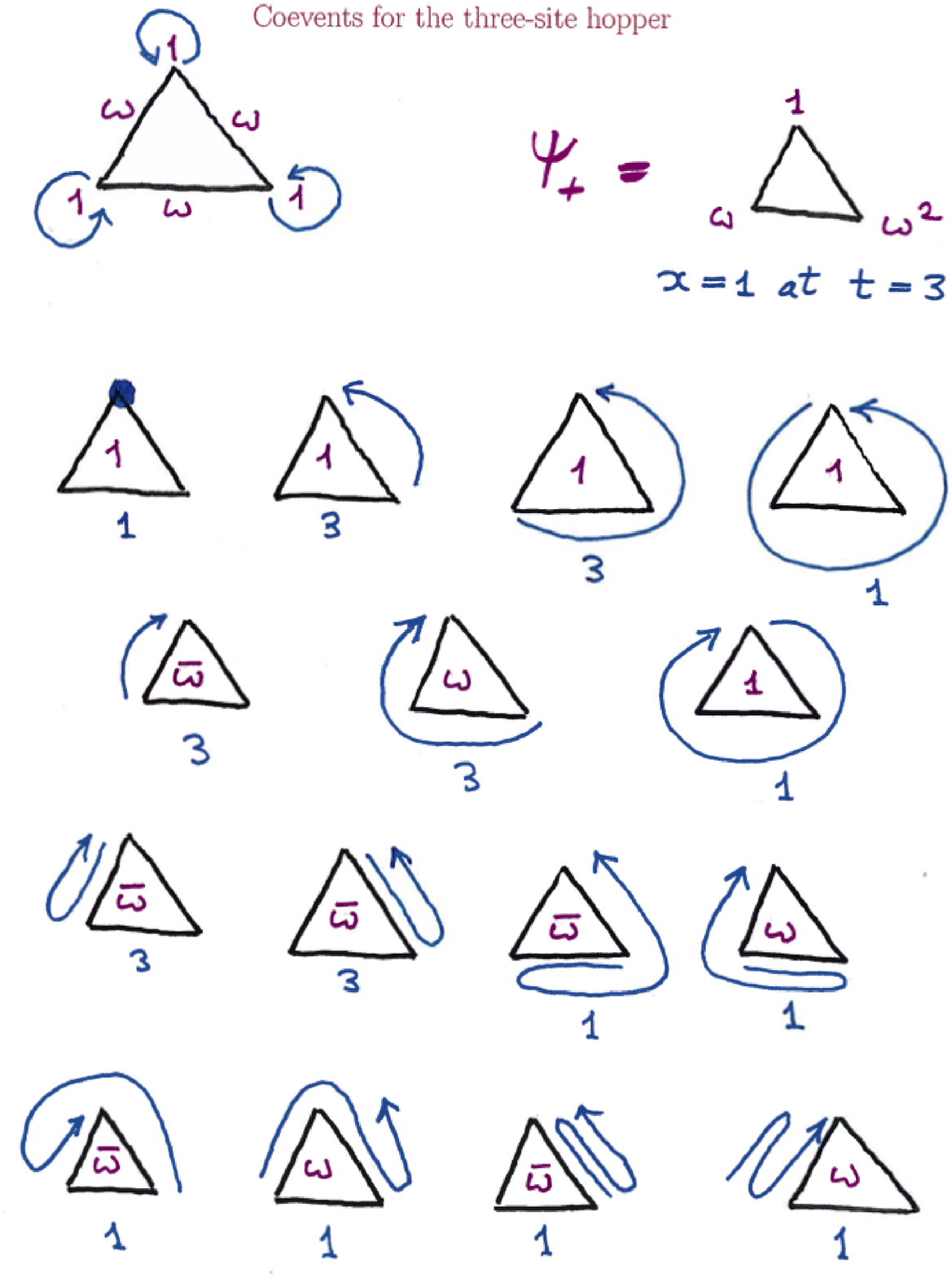} %%%
  \Caption{{\it Figure 2.}
  The 27 histories, and their amplitudes for the case of the
  ``traveling wave''.  Each path indicated by an arrow represents from 1
  to 3 possible histories differing from each other by the moments at
  which the hopper chooses to rest.  The resulting multiplicity is shown
  under the triangle, while the number inside the triangle is the amplitude
  itself.}}

\section{IV.~ Primitive coevents for the ground state and the traveling wave}           
Which combinations of the histories illustrated in figure 2
yield primitive coevents?  If we think in terms of the amplitudes
abstractly, it is easy to answer this question, and the same answer will
then apply unchanged to both cases of traveling wave and ground state.  
An
event $E$ 
will correspond to a set of amplitudes (strictly speaking a
multiset), and $E$ will be precluded precisely when the amplitudes sum
to zero.  Because the only real-linear relation among the complex numbers
$1$, $\omega$, $\omegabar$ 
is $1+\omega+\omegabar=0$, 
this will occur when, and only when, the three amplitudes occur in equal
numbers within $E$.  It is then easy to see what are the maximal
precluded multisets of amplitudes.  They are those consisting of 
6 copies each of $1$, $\omega$, and $\omegabar$. 

Now let $F$ be the support of a preclusive coevent.  In order to be
preclusive, $F$ must not fall wholly within any precluded event, and
this means that it must not be possible to adjoin further amplitudes to
those of $F$ such that the resulting multiset sums to zero.  Plainly $F$
will be protected in this way iff it contains at least 7 copies of $1$
or 7 copies of $\omegabar$.
On the other hand, we also want $F$ to be primitive, meaning minimal
among the preclusive supports.  Again it is easy to see what this means: 
it must comprise {\it precisely}   
7 copies of $1$ or 7 copies of $\omegabar$.  

In this way, we find a total of 
${12 \choose 7} + {9 \choose 7} = 828$
%% (12 choose 7) + (9 choose 7) = 828
primitive coevents of the multiplicative form $\phi=F^*$, each made up of
a total of 7 histories.  

In the case of the traveling wave, for example, one such set of
histories corresponds to the last three patterns in the first row of
figure 2.  
Interestingly, all seven of these trajectories 
move in the positive (counterclockwise) direction and 
none of them in the contrary direction.  
Thus the event, 
$P$ = ``The particle circulates exclusively in the positive sense'', 
happens in the reality described by this coevent.

Happily enough, this sense of circulation corresponds perfectly with the
``phase velocity'' of $\psi_{+}$, 
but 
we cannot assert
that all of the 828
traveling-wave coevents also affirm this same event $P$.  
In order to
quantify the tendency toward counterclockwise motion, 
then, 
let us associate a
``net circulation'' with each coevent, as the total number of
``forward'' hops less the total number of ``backward'' ones.  (So for
example the above coevent has a net circulation of $3\times 1 + 3\times
2 + 1\times 3=12$.)  Averaging this quantity over all 828 primitive
coevents yields an average net circulation of $7/23$.  A tendency toward
counterclockwise motion is therefore present but not extremely
pronounced -- just as one might expect since a lattice of only $n=3$
positions can lend no more than a very rudimentary meaning to a
derivative like ``d(phase)/d(angle)''.

The primitive preclusive coevents for the ground-state are again
supported on sets of seven histories, as we have already remarked.  By
symmetry we cannot expect a favoured sense of circulation in this case, 
but it is also interesting to ask how
``restless''
% ``peripatetic'' 
the particle proves to be.
Here, it is natural to compare with ``Bohmian'' or ``pilot wave''
conceptions of reality, since they also give meaning to the notion of
particle-trajectory.  As far as I know, 
the Bohmian ``guidance equation''
is limited to
continuum space and time, 
where the nearest analog to our hopper
ground-state might be the ground-state of a particle in a box.  Since
the phase of the Schr{\"o}dinger wave-function is independent of
position in that case, the Bohmian particle does not move at all.
Rather than explore its surroundings, it just stays put, wherever it
happens to find itself.\footnote{$^\dagger$}
{The first excited state, and indeed every standing wave of pure
 frequency, also leads to static trajectories.  In this sense a Bohmian
 particle in a box with reflecting walls would be about as far as
 possible from knowing its own energy.  However, the strictly stationary
 wave-functions are very special, and their superpositions might
 generically lead to different typical trajectories.~[12]}

It turns out that the coevents of the multiplicative scheme paint a very
different picture.  The event, ``The hopper never moves'' is of course
denied by all 828 coevents, while the contrary event that ``The hopper
never rests'' is affirmed by eight of them.  
(For such a coevent,
% $\phi=F^*$, 
all seven of the histories 
in its support
% of which $F$ is comprised 
are in
constant motion.)  Of the other 820 primitive coevents, 28 of them
affirm that the hopper 
either never rests or never moves  
(6 histories vs. 1), 
while the remaining 792 primitive supports consist
entirely of histories such that the hopper rests once and hops twice.
Moreover, for none of the coevents does the event 
``the hopper avoids some lattice-site'' 
% ``the hopper confines itself to just two of the lattice sites'' 
happen.\footnote{$^\flat$}
{Which of course doesn't mean that the complementary event, 
 ``the hopper visits all three lattice sites'' 
 ever happens either.  That's why the coevents are ``anhomomorphic'':
 with $\phi=F^*$
% and  $F^*$ 
defined as above, it can easily happen that
 two events $A$ and $B$ are complementary subsets of the space of
 histories, but both $\phi(A)$ and $\phi(B)$ vanish.}
All in all, very 
peripatetic 
indeed.

\section{V.~ {Does the hopper know its energy?}}
% \section{XXX.~ Conclusions based on this example}
%
If the multiplicative scheme (or any other coevent scheme) is a kind of
``equation of motion for coevents'', then a primitive preclusive coevent
is a kind of ``solution of the equations of motion''.
In celestial mechanics such a solution would be the orbit of a planet.
One can deduce the binding energy of the planet from a knowledge of its
orbit, but the converse is impossible since the energy is only one among
a number of orbital parameters.
Now consider some microscopic counterpart of this problem, 
like a hydrogen atom.
By analogy one should not expect to deduce a unique
coevent $\phi$ from a knowledge of the total energy,
but one might wonder whether, conversely, 
the coevent determines the energy unambiguously. 

Quantum mechanically, energies are determinate only for eigenfunctions
of the Hamiltonian operator, and the nearest analogs for these in our
hopper-world are the initial-amplitude sets, $\psi_{0}$ and $\psi_{+}$
(plus, of course, the parity-reversed set $\psi_{-}$, together with
linear combinations like the ``standing wave'', $\psi_{+}+\psi_{-}$).
We are thus led to ask whether the primitive coevents pertaining to
$\psi_{0}$ overlap with those pertaining to $\psi_{+}$.  To the extent
that the answer is negative one can indeed deduce the hopper's energy
from a knowledge of the coevent that describes its motion.  Based on the
above analysis of the primitive coevents, the required calculation is
straightforward, and the result is that the degree of overlap is zero.
No coevent is common to both $\psi_{0}$ and $\psi_{+}$.

Moreover it is possible to distinguish $\psi_{0}$ from $\psi_{+}$ in
terms of relatively elementary consequences.
For example if the hopper {never rests}
(meaning $\phi(E)=1$, where $E$ 
%% = ``the hopper never rests'' 
 is the event consisting of the fourth, seventh, and tenth through
 fifteenth histories shown in figure 2),
then $\phi$ pertains to {$\psi=\psi_0$}.
Similarly, if the hopper {moves only counterclockwise}, or 
if it {rests exactly once}, then {$\psi=\psi_+$}.

In this sense, we can say that the hopper does know its own energy.
We can also say that it ``knows its own angular momentum'', since a
similar comparison reveals that the primitive coevents pertaining to 
$\psi_{+}$ are disjoint from those pertaining to $\psi_{-}$.

Of course the example we've studied is exceptionally simple, both with
respect to the hopping amplitudes 
and the amplitudes of the ``initial states'' $\psi_{initial}$ which we
have considered.    
It would be good 
to analyze also the case of the standing wave, 
and more generally  
to extend the analysis to include
longer times, and larger lattice sizes $n$.
(The case of shorter times is also instructive.  
 For only two time-steps, one
 finds that there do exist primitive coevents common to $\psi_{+}$ and
 $\psi_{0}$.  This strengthens the impression that with the passage of
 time the hopper 
 would ``know'' more and more about $\psi_{initial}$.)
%  would ``know'' more and more about $\psi_{initial}$.~[R::petros])
%

Our hopper-world is exceptionally simple in another way too, that
relates to the radical inseparability (or ``interconnectedness'') that
seems to show up in the quantum world.  
Were we to include a second
system or
process in our idealized world, say for example a four-site hopper, 
% or even just to extend time for a few more steps, 
the coevents would change.  Obviously the global coevents would change,
but even those induced for the three-site hopper would in general be
different.  This at least is a feature of the multiplicative scheme, and
it is likely true more generally.  
It is therefore 
% especially 
% more than ordinarily 
important to study
extensions of our model of this type.

Extensions of our model in any of the directions just mentioned 
would of course be interesting.
% and they could potentially revise some of the conclusions reached so far.  
But even without them, I hope the
examples studied in this paper suffice to illustrate how one can start
to think about the quantum world without invoking the ideas of either
evolving wave-functions or external observers.

If these examples are not misleading us,
then we can already draw some conclusions of more
general validity, concerning first of all the relation between
wave-functions and descriptions of reality.  A histories-based or
``path-integral'' formulation of the sort we have been working with
has no use for the Schr{\"o}dinger equation at a fundamental level.
The basic concept of precluded event 
has a ``spacetime character'' and
refers 
directly to the histories and their amplitudes, 
% and it does not seem possible to infer preclusion from the knowledge of 
not to
any evolving
wave-function. 
On the other hand, something like a wave-function does enter into the
``initial conditions'' needed in 
setting up
% defining 
the path-integral.  In
principle one should probably replace these initial amplitudes with
cosmological boundary conditions imposed directly on the histories, but
in our hopper-world there is no moment earlier than $t=0$, and we have
assumed instead that in a more complete model, we would be able to
condense the effect of the true boundary conditions into a set of three
initial amplitudes for the hopper at $t=0$.  
This is the wave-function $\psi_{initial}$ 
in its role as ``effective summary of the past''.

What our hopper model teaches us about $\psi_{initial}$
% the resulting wave-function 
is that it is far from furnishing a detailed description of physical
reality.  Rather, the coevent which by definition does furnish such a
description is  determined by $\psi_{initial}$ only to a very limited extent.
Reality possesses far more ``internal structure'' than is reflected in a
wave-function.

This observation in turn resolves an old paradox that has recently been
emphasized in a cosmological setting by Daniel Sudarsky [13].
In its terrestrial form the paradox asks how the spherically symmetric
wave-function resulting from the decay of a spinless nucleus can be
compatible with the fact that the daughter nuclei will be found to be
localized in angle if one sets up detectors.  How does the symmetry
break?  
But in terms of coevents, there's no problem.  Just because the
$\psi$-function is symmetric, that doesn't imply the same of the
individual coevents.  Indeed, we see exactly such a phenomenon in our
hopper model, where $\psi_{initial}$ exhibits perfect rotational
symmetry while the individual coevents are completely asymmetric.  No
single coevent shares the symmetry of $\psi_{initial}$, but only the
ensemble of all $3\times828$ of them taken together.

%% are any coevents at all symmetric? no, sharp outcome!

%: Acknowledgments                           

For discussions of these questions over a long period, I would like to
thank 
  Fay Dowker, 
  Cohl Furey, 
  Yousef Ghazi-Tabatabai, 
  Stan Gudder,
  Joe Henson,
  David Rideout, 
  Sumati Surya,
  and
  Petros Wallden. 
I also thank the Foundational Questions Institute (FQXI) for a grant, 
% if i used `alisp' , which i did.
and 
the maintainers of Steel Belt Common Lisp (SBCL),
whose software I used at several points in the analysis reported
above. 
Finally, I thank the members of the Raman Research Institute (RRI) for
their warm hospitality while parts of this paper were being written.

\bigskip
\noindent
This research was supported in part by NSERC through grant RGPIN-418709-2012.
Research at Perimeter Institute for Theoretical Physics is supported in
part by the Government of Canada through NSERC and by the Province of
Ontario through MRI.

\ReferencesBegin                             

% (ref-equivalence:: )
% After the double colon put: tag-1 tag-2 tag-3 ...

\ref [1] % papers defining or using n-site hopper
Rafael D.~Sorkin,
``Toward a `fundamental theorem of quantal measure theory'$\,$''
 \journaldata{Mathematical Structures in Computer Science}{22(5)}{816-852}{2012} \lbr
 \eprint{http://arxiv.org/abs/1104.0997} \lbr
 \eprint{http://www.pitp.ca/personal/rsorkin/some.papers/141.fthqmt.pdf}.

\ref [2]
S.~Gudder and Rafael D.~Sorkin,
``Two-site quantum random walk''
 \journaldata{General Relativity and Gravitation}{43}{3451-3475}{2011}
 \arxiv{1105.0705}  \lbr
 \eprint{http://www.pitp.ca/personal/rsorkin/some.papers/}

\ref [3] % refs to coevents and to MSk
Rafael D. Sorkin,
``Quantum dynamics without the wave function''
 \journaldata{J. Phys. A: Math. Theor.}{40}{3207-3221}{2007}
 (http://stacks.iop.org/1751-8121/40/3207)
\eprint{quant-ph/0610204} 
\eprint{http://www.pitp.ca/personal/rsorkin/some.papers/} 

\ref [4]
Petros Wallden,
``The coevent formulation of quantum theory'', \lbr
 \eprint{http://arxiv.org/abs/1301.5704}

\ref [5]
  Rafael D.~Sorkin,
  ``Logic is to the quantum as geometry is to gravity''
 in G.F.R. Ellis, J. Murugan and A. Weltman (eds),
 {\it Foundations of Space and Time: Reflections on Quantum Gravity} 
 (Cambridge University Press, 2012) \lbr
   \arxiv{arXiv:1004.1226 [quant-ph]} \lbr
   \eprint{http://www.pitp.ca/personal/rsorkin/some.papers/}

% \ref [R::coevent-review] petros review of coevents or some other review
% \ref [R::capetown] good to cite capetown for more detail

\ref [6] % def of mu, qmt in general 
Rafael D.~Sorkin,
``Quantum Mechanics as Quantum Measure Theory'',
   \journaldata{Mod. Phys. Lett.~A}{9 {\rm (No.~33)}}{3119-3127}{1994}
   \eprint{gr-qc/9401003} \lbr
   \eprint{http://www.pitp.ca/personal/rsorkin/some.papers/80.qmqmt.pdf}

\ref [7]
Roberto B.~Salgado, ``Some Identities for the Quantum Measure and its Generalizations'',
 \journaldata{Mod. Phys. Lett.}{A17}{711-728}{2002}
 \eprint{gr-qc/9903015}

\ref [8]
Sumati Surya and Petros Wallden, ``Quantum Covers in Quantum Measure Theory'',
\arxiv{0809.1951}
\journaldata{Foundations of Physics}{40}{585-606}{2010}

% \ref [R::born-rule] [[papers showing how macroscopic amplification, decoherence etc leads to born rule]]

\ref [9]  % more on MSk and its consequences (plus next two)
Yousef Ghazi-Tabatabai and Petros Wallden, 
``Dynamics \& Predictions in the Co-Event Interpretation'', \lbr 
 \journaldata{J. Phys. A: Math. Theor.} {42}{235303}{2009}
 \arxiv{0901.3675}

\ref [10]
Yousef Ghazi-Tabatabai and Petros Wallden, 
``The emergence of probabilities in anhomomorphic logic'', 
 \journaldata{Journal of Physics: Conf. Ser.}{174}{012054}{2009} \lbr
 \arxiv{0907.0754 (quant-ph)}

\ref [11]
Fay Dowker and Yousef Ghazi-Tabatabai, \lbr
``The Kochen-Specker Theorem Revisited in Quantum Measure Theory'',
 \journaldata {J.Phys.A}{41}{105301}{2008}
 \arxiv{0711.0894 (quant-ph)}

%% \ref [R::precl-sep] [[proof of precl sep lemma in MSk (whence sharp final position)]]

\ref [12] Lucien Hardy, remark at a conference.

%% \ref [R::petros] petros conjecture that with time the hopper knows psi fully

\ref [13] % paradox of structure formation
 Daniel Sudarsky,
 ``Shortcomings in the Understanding of Why Cosmological Perturbations Look Classical''
 \journaldata{Int. J. Mod. Phys.}{D20}{509-552}{2011}
  arXiv:0906.0315 [gr-qc]

\end                                         

%: Outline mode stuff (put here so doesn't need to be commented out) 

(prog1 'now-outlining
  (Outline* 
     "\f"                   1
      "%------"             1
      "%:  "                2
      "%: "                 1       
      "\\Abstract"          1
      "\\section"           1
      "\\subsection"        2
      "\\ReferencesBegin"   1
      "% (ref-equivalence"  2
      "\\ref "              2
      "\\end